\def\be{\begin{equation}}
\def\ee{\end{equation}}
\def\ba{\begin{eqnarray}}
\def\ea{\end{eqnarray}}
\def\l{\label}

\documentstyle[12pt,epsf,epsfig,rotating,a4,subfigure]{article} \begin{document} 

\title{Mass dependence of HBT correlations in $e^+e^-$ annihilation }

\author{A.Bialas, M.Kucharczyk, H.Palka and K.Zalewski \\ M.Smoluchowski
Institute of Physics \\Jagellonian University, Cracow\thanks{Address:
Reymonta 4, 30-059 Krakow, Poland;}\\ Institute of Nuclear Physics,
Cracow\thanks{Address: Kawiory 26a, 30-055 Krakow, Poland}}

 \maketitle

\begin{abstract}

Mass dependence of the effective source radii, observed in hadronic
$Z^0$ decays by several LEP I experiments, is analyzed in a model which
assumes proportionality between four-momentum of a produced particle and
the four-vector describing its space-time position at the freeze-out. It
is shown that this relation (commonly accepted in description of
high-energy collisions) can explain the data, provided all particles are
emitted from a "tube" of $\sim 1$ fm in diameter at a constant proper
time $\sim 1.5$ fm. 

\end{abstract}
 
\section {Introduction}

Recently, two of us have pointed out \cite{bz1} that the generalized
Bjorken-Gottfried hypothesis \cite{bg}, relating the space-time position
of a hadron produced in a high-energy collision to its 4-momentum, can
qualitatively explain the mass-dependence of the interaction radii
observed in $e^+e^-$ annihilation at LEP I \cite{LEP_2d,LEP_1d}. 
As discussed in
detail in \cite{bz1}, this effect is a manifestation of the
well-established observation \cite{bow} that a correlation between the
momentum and the emission point of a particle can drastically affect the
results of the HBT correlation experiment. In the present note we want
to explore in more detail the idea formulated in \cite{bz1} in order to
verify, if it indeed provides a viable framework for the understanding
of the mass dependence of the HBT parameters.

The generalized Bjorken-Gottfried hypothesis, as formulated in \cite{bz1},
postulates the linear relation between the 4-momentum of the produced
particle and the space-time position at which it is produced\footnote{To
our knowledge, the first application of  relation (\ref{1}) to a discussion of
the quantum interference between  identical particles was proposed
(in a different context) by Csorgo and Zimanyi \cite{cz}}:
\begin{equation}
q_{\mu}= \lambda x_{\mu}. \label{1}
\end{equation}
Relation (\ref{1}) implies
\begin{equation}
\lambda = \frac{M_{\perp}}{\tau}   \label{3}
\end{equation}
where $M_{\perp}^2 = E^2-q_{\parallel}^2 = m^2 +q_{\perp}^2$, and 
\begin{equation}
\tau= \sqrt{t^2-z^2}    \label{2}
\end{equation}
is the {\it longitudinal} proper time  after the collision ($t$
 and $z$ are time and longitudinal position of the production point).

Since this picture is purely classical, it represents only a
qualitative
idea, whose application to the description of the actual data requires an
adequate reformulation  taking into account the  effects of the quantum
nature of the system considered. In \cite{bz1} we have proposed to use 
(\ref{1}) and (\ref{3}) as a guide-line for construction of the
source function\footnote{It was called  there a "generalized Wigner function".}
 $S(P,X)$ \cite{wh,all} related to the density matrix in momentum
space by a Fourier transform
\begin{eqnarray}
\rho(q=P+\frac12Q,q'=P-\frac12Q)=
 \int d^4X e^{iQX} S(P,X). \label{5}
\end{eqnarray}
All variables are four dimensional, so that both space
and time integrations are involved.
Thus specifying  the source function  fixes completely the
single particle properties of the model.

Similarly as the standard Wigner function \cite{wig}, $S(P,X)$ gives
approximately (as far as possible without contradicting quantum
mechanics) the single-particle distribution in momentum and in
space-time. Therefore it has an intuitive meaning\footnote{It should be
realized that, in contrast to the standard Wigner function which relates
the particle wave functions at different positions but at {\it the same
time}, the {\it } source function relates the particle
production amplitudes at different positions and {\it at different
times.} (as is clearly seen from (\ref{5})). Consequently some care is
needed in order to assess its physical interpretation.}.
which can be exploited for implementation of the relations
(\ref{1})-(\ref{3}).

The construction  of $S(P,X)$  (described in the next section)
 requires specification of several details
which can only be determined from a confrontation with data. We thus
analyse the single- and two-pion distributions in the framework of the
scheme proposed in \cite{bz1} and compare them with the data from the DELPHI
experiment at LEP I \cite{pt2note} in order  (i) to pin down the parameters
describing the system and (ii) to verify that their values are
reasonable, i.e., that the whole scheme is not unrealistic.
Our conclusion is rather optimistic:  the proposed scheme seems to
describe adequately the LEP I data.
As, in addition,  it provides a reasonable intuitive picture of 
the hadronization, one may  hope that it represents a
useful framework for the description of the general properties of this process.

In the next two sections we describe the construction of the 
source function and calculate the single particle density matrix in
momentum space. In Section 4 the formulae for two-particle correlation
functions are derived. Comparison with the data is discussed in Section
5 and 6. The last section  contains our conclusions and outlook.

\section{The  Source Function }

To implement the conditions (\ref{1})-(\ref{3}) we postulate the  
source function  in the factorized form
\begin{equation}
S(P,X) =  F(\tau) S_{\parallel} S_{\perp}  \label{6}
\end{equation}
where 
\begin{equation}
S_{\parallel}= 
\exp\left[\frac1{2\delta_{\parallel}^2}\left(P_+-\frac{M_{\perp}}{\tau}
 X_+)\right) \left(P_--\frac{M_{\perp}}{\tau} X_-)\right)\right]
\label{7}
\end{equation}
and  
\begin{eqnarray}
 S_{\perp}=
 \exp\left[-\frac{X_{\perp}^2}{2r_{\perp}^2}
\right] 
\exp\left[-\frac{\left(\vec{P}_{\perp}-
\frac{M_{\perp}}{\tau}\vec{X}_{\perp}\right)^2}
{2\delta_{\perp}^2}\right].     \label{3c} 
\end{eqnarray}
Here
\begin{equation}
X_{\pm} = t\pm z;\;\;\;\; P_{\pm}= P_0\pm P_z.    \label{3a}
\end{equation}
so that 
\begin{equation}
M_{\perp}^2=P_+P_-;\;\;\;\;\; \tau^2=X_+X_-. \label{3b}
\end{equation}
We have used Gaussian forms in order  to simplify the 
evaluation of the Fourier transform (\ref{5}). 
As shown below, this admittedly crude 
assumption seems sufficient at the present level of analysis and accuracy
of the data.

The first exponential factor in $S_{\perp}$ represents a standard\footnote{
For  related approaches see, e.g., \cite{cs}.}
cylindrically symmetric "tube" of radius $r_{\perp}$ in
configuration space\footnote{To simplify the argument, we ignore the
rapidity and $z$ dependence of the single particle
spectrum. This seems a reasonable approximation in the central rapidity
region at high energy and can be easily removed, if necessary.}. The
remaining exponential introduces a correlation between the momentum and
the emission point of the particle, as required by the generalized
Bjorken-Gottfried condition (\ref{1}). Such correlations are known to
influence strongly the HBT effect on particle spectra \cite{bow} and
were shown in \cite{bz1} to induce the mass dependence of the HBT radii.
The reason for the mass dependence is easily seen from (\ref{3}): at
fixed $\tau$, the correlation between momentum and position depends on
transverse mass.

The parameters $\delta_{\parallel}$ and $\delta_{\perp}$ parametrize the
correlation lengths or - in other words - the size of the region
(centered at $X_{\mu}$) from which the observed particles emerge. They
are arbitrary, apart from inequalities required for consistency with the
quantum uncertainty principle, see, e.g., \cite{bz2}, which put on them
some weak lower limits.

Finally, the function $F(\tau)$ gives the distribution of the proper time
$\tau$ at which the particles are created.

Using now the formula (\ref{5}) and the formulae of this section one can
evaluate the single-particle density matrix in momentum space, which is
a necessary step in the evaluation of both the single-particle spectrum and
the two-particle HBT correlations. This calculation is outlined in the next
section.

\section{ Density matrix in momentum space}

Substituting (\ref{6}) into (\ref{5}) we have
\begin{equation}
\rho(q,q') = \int \tau d\tau  F(\tau) \rho_{\parallel} \rho_{\perp}  \label{3.1}
\end{equation}
where
\begin{equation}
\rho_{\perp} = \int d^2X_{\perp}
 S_{\perp} e^{-i\vec{X}_{\perp}\vec{Q}_{\perp}}
\label{3.2}
\end{equation}
and 
\begin{equation}
\rho_{\parallel}= \int d\eta e^V. \label{3.3a}
\end{equation}
Here $\eta$ is the pseudorapidity:
\be
\eta= \log\left(\frac{X_+}{X_-}   \right)     \label{x3.3}
\ee
and
\begin{equation}
V=\frac1{2\delta_{\parallel}^2}\left(P_+-\frac{M_{\perp}}{\tau
} X_+)\right) \left(P_--\frac{M_{\perp}}{\tau} X_-)\right)
+i(Q_0t-Q_{\parallel}z).  \label{3.3}
\end{equation}
From (\ref{5}) we also see that
\begin{equation}
P=\frac12(q+q');\;\;\;\; Q= q-q'   \label{3.3b}
\end{equation}
The Gaussian integral (\ref{3.2}) for $\rho_{\perp}$ 
can be explicitly performed with the result
\begin{equation}
\rho_{\perp}(\vec{q}_{\perp},\vec{q}_{\perp}')= 
2\pi r_{eff}^2
\exp \left( -\frac{\vec{P}^2}{2\omega^2}-\frac{\vec{Q}_{\perp}^2r_{eff}^2}2 \right)
\exp\left[-i\frac{M_{\perp}\tau v^2}{\omega^2} 
\vec{P}_{\perp}\vec{Q}_{\perp}\right]    
\label{3.4}
\end{equation}
where 
\begin{equation}
\omega^2= M_{\perp}^2 v^2 +\delta_{\perp}^2;\;\;\;\ v^2= r_{\perp}^2/\tau^2 
;\;\;\; r_{eff}^2= \frac{r_{\perp}^2\delta_{\perp}^2}{\omega^2}       \label{3.6}
\end{equation}
The longitudinal integral is somewhat more complicated. We first
express the energies and longitudinal momenta in terms of the respective rapidities
\begin{equation}
P_{\pm} = M_{\perp} e^{\pm Y};\;\;\; Q_0=m_{\perp} \cosh y -m_{\perp}'
\cosh y';\;\;\; Q_{\parallel}= m_{\perp} \sinh y -m_{\perp}' \sinh y'
\label{3.7}
\end{equation}
Substituting this into (\ref{3.3}) we obtain after some algebra
\begin{eqnarray}
V=\frac { M_{\perp}^2}{\delta_{\parallel}^2}(1-\cosh(Y-\eta))
\nonumber \\+i\frac{\tau}{M_{\perp}}\left(\vec{P}_{\perp}\vec{Q}_{\perp}
 \cosh(Y-\eta)
+ m_{\perp}m'_{\perp}\sinh(y-y') \sinh(\eta-Y) \right)  \label{3.7a}
\end{eqnarray}
One sees from this formula that V
depends only on $Y-\eta$ and $y-y'$. Consequently, after integration
over $\eta$ the result depends only on $y-y'$.

The integral (\ref{3.3a}) can now be evaluated. We first change the
variable $\eta -Y \rightarrow \xi$. Then we express the hyperbolic
functions of $\xi$ by exponentials. Using the integral representation
\begin{equation}
K_0(z)= \int_0^{\infty} e^{-z \cosh y} dy   \label{3.8}
\end{equation}
we obtain
\begin{equation}
\rho_{\parallel} =2 \exp\left( \frac{M_{\perp}^2}{\delta_{\parallel}^2}\right)
K_0(s)      \label{3.9}
\end{equation}
with 
\begin{equation}
s^2 =\frac{M_{\perp}^4}{\delta_{\parallel}^4}-\frac{\tau^2}
{M_{\perp}^2}(\vec{P}_{\perp}\vec{Q}_{\perp})^2
-2i\frac{\tau M_{\perp}}{\delta_{\parallel}^2}
\vec{P}_{\perp}\vec{Q}_{\perp} +\frac{\tau^2}{4M_{\perp}^2} m_{\perp}^2m_{\perp}'^2\sinh^2(y-y')
 \label{3.10}
\end{equation}
Using the identities
\begin{eqnarray}
m_{\perp}\cosh y +m'_{\perp}\cosh y'=2M_{\perp}\cosh Y \nonumber \\
m_{\perp}\sinh y +m'_{\perp}\sinh y'=2M_{\perp}\sinh Y  \label{3.11}
\end{eqnarray}
we arrive, after some algebra, at our final formula
\begin{eqnarray}
s^2=\frac {M_{\perp}^4}{\delta_{\parallel}^4}
-\tau^2 Q_t^2 
-i\frac{\tau M_{\perp}}{\delta_{\parallel}^2}(m_{\perp}^2 - m_{\perp}'^2)
\label{3.12}
\end{eqnarray}
where
\be
Q_t^2= Q_0^2-Q_{\parallel}^2      \label{3.13}
\ee
This completes the evaluation of the single-particle density matrix.

\section {Single-particle distribution and two-particle correlation function}

The single particle distribution is given by the diagonal elements of the
density matrix. From the formulae of the previous section we thus obtain


\begin{equation}
\rho(q)\equiv\frac {dn}{dy d^2q_{\perp}}= 2\pi r_{\perp}^2 \delta_{\perp}^2
 \exp\left(\frac{m_{\perp}^2}{\delta_{\parallel}^2}\right)
K_0\left(\frac{m_{\perp}^2}{\delta_{\parallel}^2}\right) I(q_{\perp}^2)
\label{4.1}
\end{equation}
where
\begin{equation}
I(q_{\perp}^2) = \int \tau d\tau F(\tau) \bar{ \omega}^{-2} 
\exp\left(-\frac{ q_{\perp}^2}{2\bar{\omega}^2 })\right)
\label{4.2}
\end{equation}
and
\begin{equation}
\bar{\omega}^2= m_{\perp}^2 v^2 + \delta_{\perp}^2  \label{4.3}
\end{equation}

To obtain information on the two-particle correlation function, one has
to make further assumptions. We follow the standard treatment
\cite{wh,all,bz2}, assuming that one
can evaluate the two-particle correlation function as if
there were  no other correlations between
 particles except for those induced by quantum interference. Under this
condition the normalized two-particle correlation function is given by
\begin{equation}
C(q_1,q_2)= \frac{|\rho(q_1,q_2)|^2}{\rho(q_1)\rho(q_2)}    \label{4.4}
\end{equation}
The final point we want to discuss is the selection of the variables used
for comparison with the data, and the corresponding Jacobians.

The system of two identical particles is fully described by the 6
components of the momenta. The phase-space volume is
\be
d\Omega = dy dy' d^2q_{\perp} d^2q'_{\perp}=2\pi dydy'dP_{\perp}^2
d Q_{\perp}^2 d\phi   \l{4.5}
\ee
where $\phi$ is the angle between $\vec{P}_{\perp}$ and
$\vec{Q}_{\perp}$.

As seen from the formulae of Section 3, it is  convenient to replace the two rapidities
by $Q_t^2$ and $Y$. Using the identity
\be
 Q_t^2= (m_{1\perp}-m_{2\perp})^2 -4 m_{1\perp}m_{2\perp} \sinh^2
\left(\frac{y_1-y_2}{2}\right)   \label{4.6}
\ee
one obtains
\be
dy_1dy_2 =\frac{dY dQ_t^2}{\sqrt{Q_t^4
-2(m_{1\perp}^2+m_{2\perp}^2)Q_t^2+( m_{1\perp}^2-m_{2\perp}^2)^2}}.
\l{4.7}
\ee
Thus everything can be expressed in terms of
$Y,Q_t^2,Q_{\perp}^2,P_{\perp}^2$ and $\phi$. It is sometimes
convenient, however, to consider other variables. For completeness we 
give below some kinematic relations which can be useful in data
analysis.
\be
\mu^2\equiv 2(m_{1\perp}^2 +m_{2\perp}^2) =4 m^2+ 4P_{\perp}^2
+Q_{\perp}^2      \l{4.8}
\ee
\be
Q_{out} = |Q_{\perp}| \cos \phi ;\;\;\;\; M_{\perp}^2 =\mu^2 -Q_t^2  \l{4.9}    
\ee
\be
m_{1\perp}^2-m_{2\perp}^2= 2 |P_{\perp}||Q_{\perp}|\cos \phi =2|P_{\perp}|
Q_{out}
\l{4.10}
\ee

\section{ Data: Single-particle distribution}

Eq.(\ref{3.1}) represents the density matrix as an integral over the
proper time $\tau$ at which the particles are produced. In the present
paper, following \cite{bz1}, we shall accept the approximation that the
production happens in a very narrow interval of $\tau$, so that the
integration over $\tau$  simply amounts to introduce a fixed 
value $\tau=\tau_0$ in the formulae of the previous section. In this way
the unknown function $F(\tau)$ is replaced by one parameter, $\tau_0$.
The other three parameters are: $v$, $\delta_{\perp}$ and $
\delta_{\parallel}$, each with a very clear physical meaning.

The first step in the data analysis should be the correct description of
the single particle spectrum using the
Eq.(\ref{4.1}). This also allows to restrict somewhat the values of the
four parameters we have to our disposal.

For this purpose the data sample of $\approx$ $3*10^{5}$ $Z^0$ hadronic decays
from the DELPHI experiment was used \cite{pt2note}. Hadronic events
have been selected using standard DELPHI cuts \cite{Hamacher} which limit
the contamination of beam-gas,$\gamma\gamma$,$\mu\mu$ and ee events to less
than 0.1\% and $\tau\tau$ contamination to below $0.2\%$.
In this study only two-jet like events were considered (Thrust $\ge$ 0.95).
The distribution of  transverse momentum  with respect to the event thrust 
axis was constructed for tracks which are not obvious decay products of $V^{0}$'s 
and are not positively identified as K's or p's. 

The formula (\ref{4.1}) has been fitted to the distribution in the
$q_{\perp}^{2}$ region up to 2.0 $GeV^{2}$ which contains $99\%$ of the
data. The 
small $q_{\perp}^{2}$ \( < 0.020 GeV^{2}\) interval was excluded from
fits because it is particularly uncertain experimentally
due to the contribution of low momentum pions
from the $D^{*}$ decays and to a possible distortion of the 
distribution by imperfect approximation of
the  primary quark direction by the measured thrust axis.

The unknown parameters ($v$, $\delta_{\perp}$ and $\delta_{\parallel}$ plus
normalization) were determined by minimizing a $\chi^{2}$
function with the MINUIT program \cite{minuit}. 
The parameter $\tau_{0}$ was kept fixed in fits  at the value of  1.5  fm.
Since correlations between some fit parameters are high, 
we used a 3 standard 
deviation correlation ellipse for each parameter pair to  estimate their 
statistical errors.
In addition, the stability of the results of the fit has been checked by varying
the range of fitted $q_{\perp}^{2}$.
The extreme values of a parameter resulting from these fits were 
used as a measure of the range of a parameter allowed by the data.

\begin{figure}[ht]
\begin{center}
\setlength{\unitlength}{1mm}
\begin{picture}(130,130)
\epsfxsize=130mm
\epsfysize=130mm
\epsffile{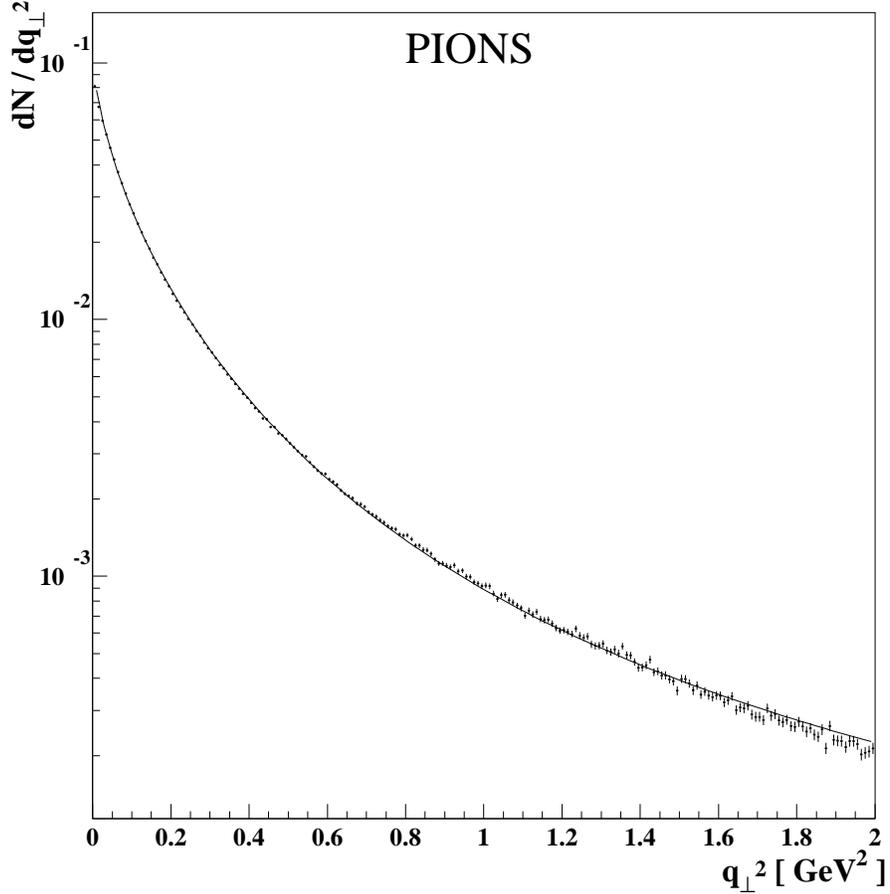}
\end{picture}
\end{center}
\caption{  The fit of the formula (\ref{4.1}) to $q_{\perp}^{2}$ distribution of pions,
           (${\it v}=0.286$ , $\delta_{\perp}=0.303$ , 
           $\delta_{\parallel}=0.172$) }
\label{pt2fit}
\end{figure}

Although the fits converged easily to a deep minimum, 
their $\chi^{2}$ value was not always statistically acceptable.
Nevertheless, we accept these fits as long as the shape of the experimental
distribution is correctly reproduced over a wide range of $q_{\perp}^{2}$. 
The too high  $\chi^{2}_{min}/dof$ value is attributed to underestimated 
errors (imperfect treatment of systematics, no background subtraction etc) 
rather, than to failure of the fitted hypothesis.
An example fit is shown in Fig.\ref{pt2fit}.
The fit results are summarized in Table~\ref{parameters}.

\begin{table}[ht]
\caption[]
        {Model parameters determined in the fit. }
$$\begin{tabular}{|c|c|c|c|}
\hline
                          &                            &
 & \\
        {\it v}           &  $\delta_{\perp}$( GeV )   &$\delta_{\parallel}$( GeV ) 
 &$\chi^{2}/dof$\\
                          &                            &
 & \\
\hline
\hline
                           &                           &
 & \\
 $0.286^{+0.014}_{-0.010}$ & $0.303^{+0.010}_{-0.021}$ & $0.172^{+0.178}_{-0.015}$
 & 397/193      \\
                           &                           &
 &  \\
\hline
\end{tabular}$$
\label{parameters}
\end{table}

\section{Two-particle correlation function}   

\begin{figure}[ht]
         \begin{center}
          \begin{tabular}[t]{cc}
           \subfigure[]{\epsfig{figure=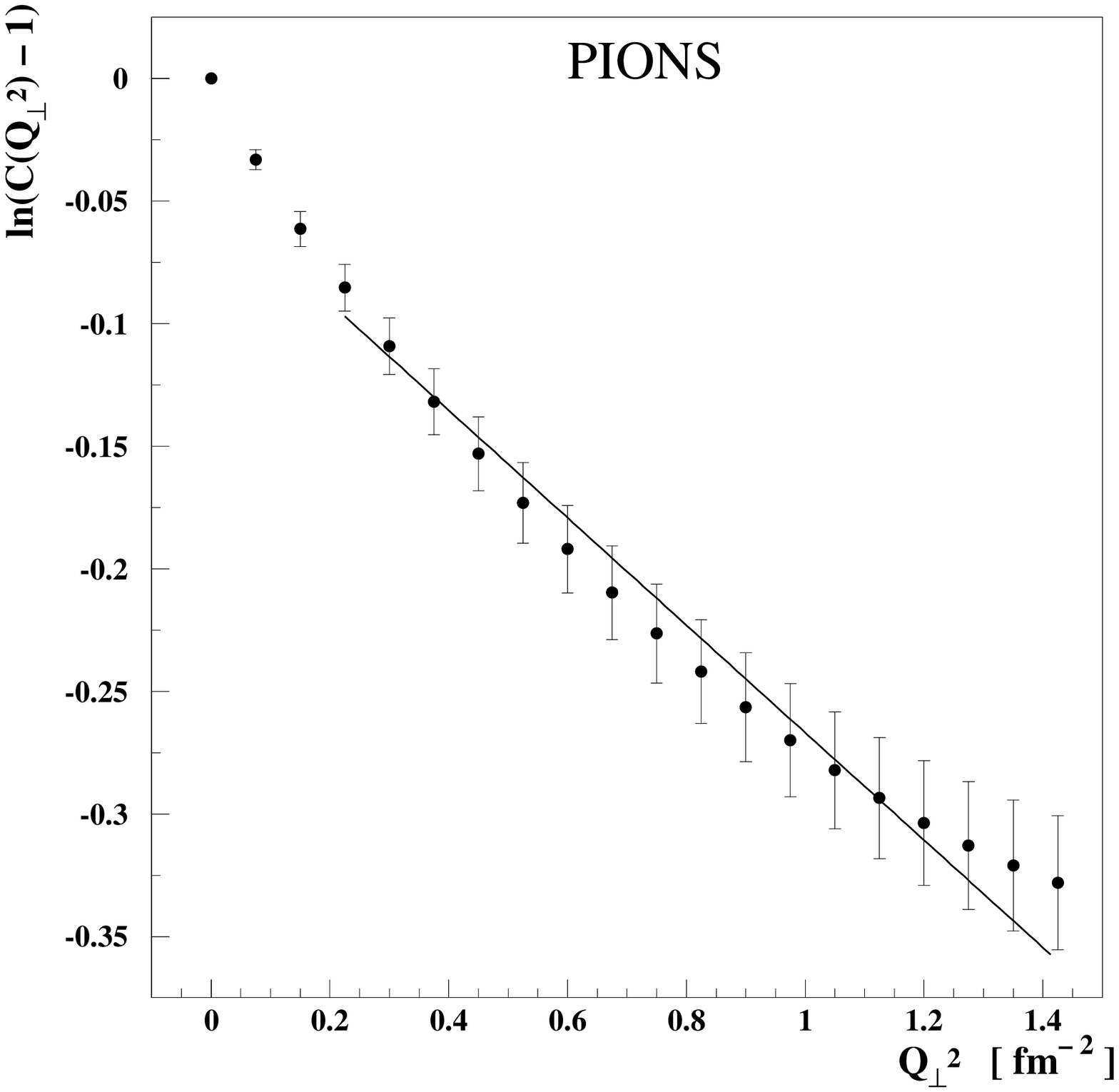,width=0.5\linewidth}}
           \subfigure[]{\epsfig{figure=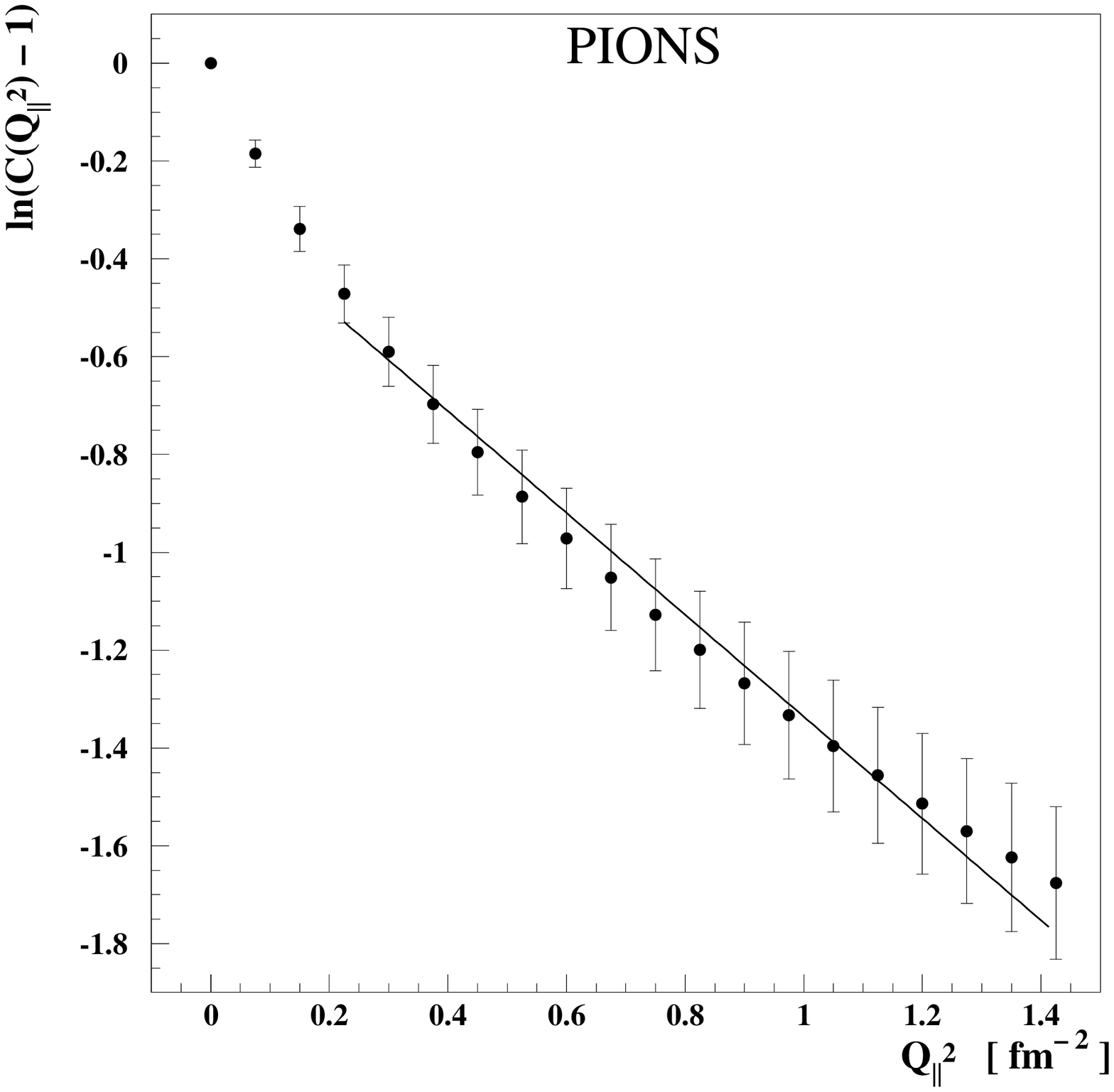,width=0.5\linewidth}}
          \end{tabular}
         \end{center}
         \caption{Correlation function vs $Q_{\perp}^{2}$ (a) and 
                  $Q_{\parallel}^{2}$ (b) for $\pi$}  
         \label{correlator}
\end{figure}

The two-particle correlation function is
calculated according to formula (\ref{4.4}) using the values of the 
parameters from Table~\ref{parameters}. Correlations between parameters
determined by the fit are taken into account by sampling the area of the
3$\sigma$ correlation ellipse for each parameter pair
with a Gaussian distribution. 
At the presence of correlations between the parameters this procedure results in a
much more reliable estimate of the statistical error on the calculated
correlation function value than the  1-dim. errors given by MINUIT.
Systematic effects of the fit are accounted for by repeating
calculations for extreme values of the parameters allowed by the
$q_{\perp}^{2}$ distribution. This procedure was needed especially for
the parameter $\delta_{\parallel}^{2}$ which is the least constrained by the
one-particle distribution. In this way the value of the correlation
function for each particle was determined together with its error, where
both error contributions described above have been added linearly. The
result of the calculation for pions is shown in Figs.\ref{correlator}a,b 
for the  perpendicular and  parallel components,
respectively. 

\begin{figure}[ht]
         \begin{center}
          \begin{tabular}[t]{cc}
           \subfigure[]{\epsfig{figure=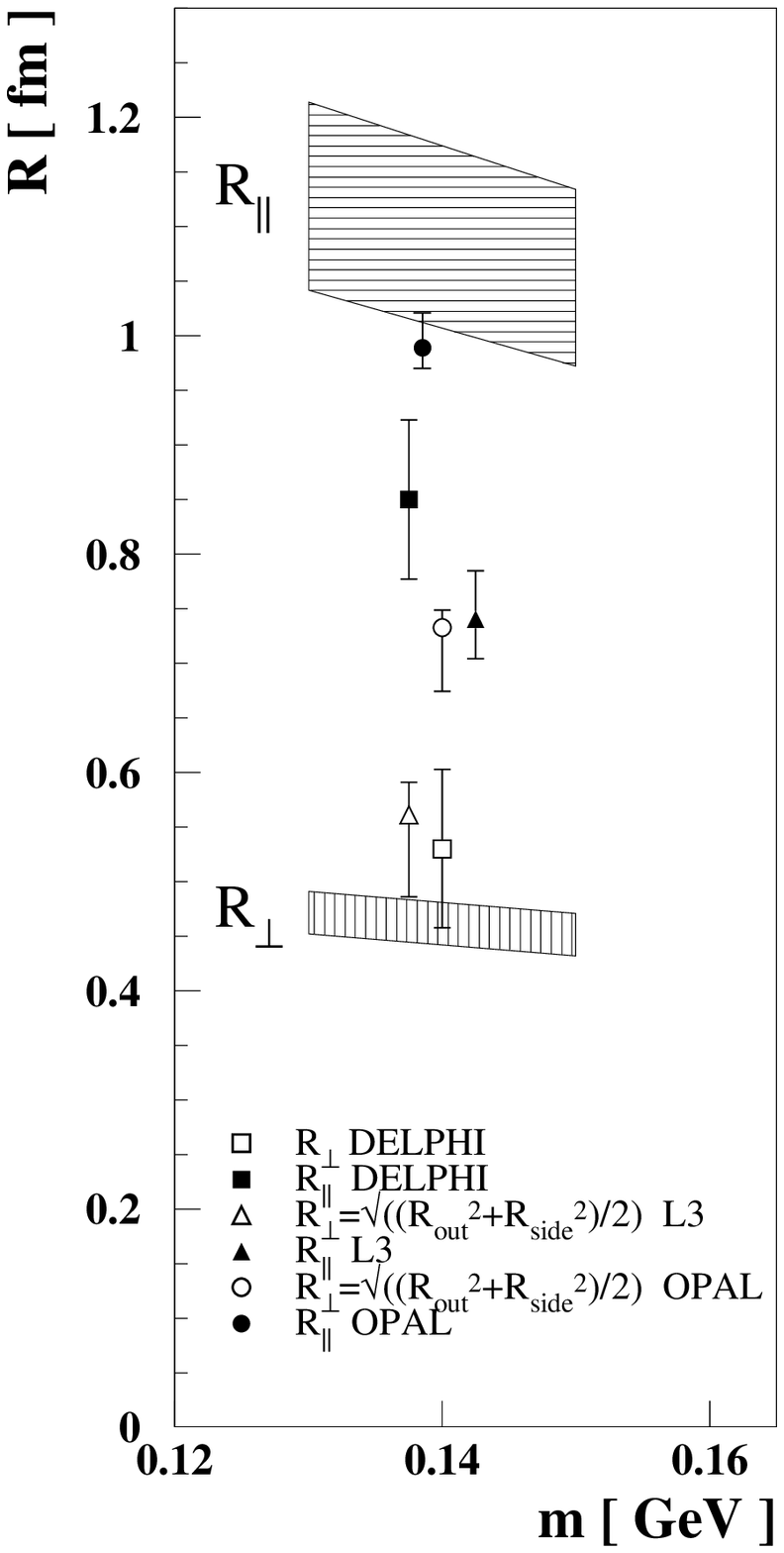,width=0.4\linewidth,height=120mm}}
           \subfigure[]{\epsfig{figure=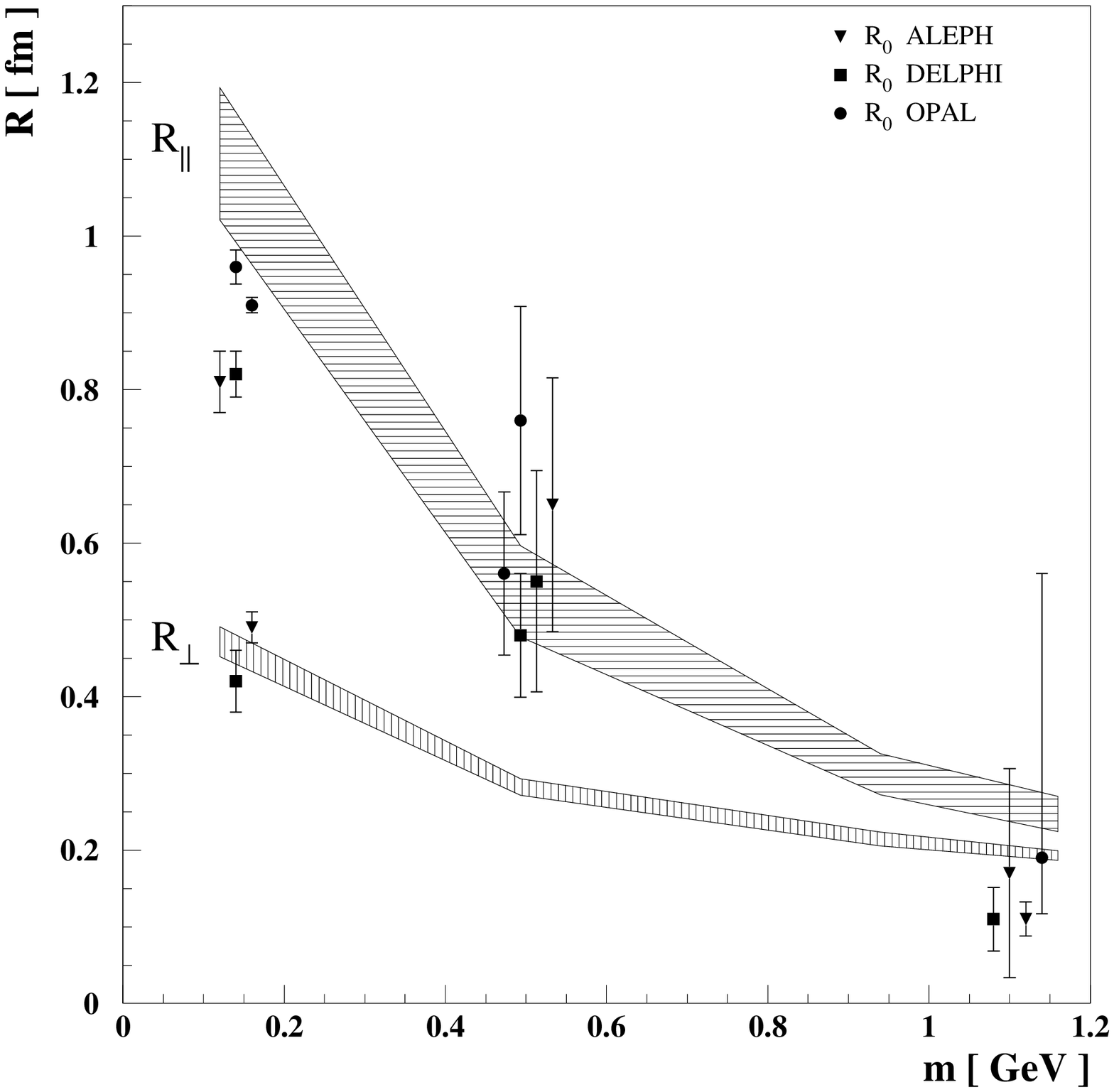,width=0.6\linewidth,height=120mm}}
          \end{tabular}
         \end{center}
         \caption{ $R_{\parallel}$ and $R_{\perp}$ calculated from the model 
                  (shaded bands) : (a) for $\pi$, (b) for $\pi$,K,p,$\Lambda$. 
                  Data points in (a) represent results of 2- and 3-dimensional
                  analyses of LEP data \cite{LEP_2d}. Data points in (b) represent
                  1-dim source radius $R_{0}$ \cite{LEP_1d} }  \label{R_massdep}

\end{figure}

One sees that these functions are not gaussian.
Nevertheless, to compare with the existing data, we have
approximated the results by Gaussians in the region $0.100 GeV < Q <
0.250 GeV$ (Fig.~\ref{correlator}). 
The exclusion of the small $Q^{2}$ region may be justified
by the well-know fact that the measurements of the correlation function
at very small $Q^{2}$ are very uncertain (and often this region is
omitted in the data analyses at LEP). 
The results of this procedure for pions  are shown in Fig.~\ref{R_massdep}(a) 
together with  measurements of the three LEP experiments \cite{LEP_2d} 
done in correlation studies in two- and three dimensions. 

In the experimental studies
 $Q^2$ is decomposed into three  components in the
longitudinal center-of-mass  (LCMS) frame,
where the sum of the three-vector momenta is perpendicular to the thrust
axis. Choosing $Q_{L}$ parallel to the thrust axis, 
$Q_{out}$ parallel
to the sum of the momenta (in the LCMS frame) 
and $Q_{side}$ orthogonal to both, the decomposition
reads:
\be
 Q^{2} = Q_{L}^{2} + Q_{side}^{2} +Q_{out}^{2}(1-\beta^{2})   ;
\;\;\;\;\; \beta = \frac{p_{out}^{1} + p_{out}^{2}}{E_{1} + E_{2}}
\ee
DELPHI performed the analysis in two dimensions, determining correlation
radii corresponding to  $Q_{T}^{2} = Q_{side}^{2} +Q_{out}^{2}$ ( $R_{\perp}$ )
and to $Q_{L}^{2}$ $(R_{\parallel})$. The 
L3 and OPAL analyses where done in three components, therefore, as the measure
of $R_{\perp}$ we have chosen 
$R_{\perp} = \sqrt{\frac{R_{out}^{2} + R_{side}^{2}}{2}}$.
The spread of the experimental results is considerable, which might be due
to different methodology (eg. different reference sample, corrections for
Coulomb repulsion etc) and differing phase space regions selected. 
However the  clustering of
the data points for $R_{\parallel}$ at larger values than those
for  $R_{\perp}$ is clearly visible.
Shaded  bands in the plot represent the errors on the calculated
$R_{\perp}$ and $R_{\parallel}$.

The most remarkable feature seen in this figure is that the model
predicts, in agreement with experiment, 
 a longitudinal radius much larger than the transverse one. One
also sees that both radii fall in the ballpark of about $1$ fm which is
hardly surprising. It is also clear that -within the rather large
theoretical and experimental errors- the model seems to account for
the gross features of the data.

The calculations were also done for kaons, protons and $\Lambda$'s. 
Since the available data samples for these particles are much smaller
than those for pions,
their $q_{\perp}^{2}$ distributions are not discriminative enough 
to  pin-point reliably the model parameters. Therefore the parameters
were taken the same as for pions (given in Table~\ref{parameters}).
This assumption is to be verified 
 once  better data are available but we have checked that it
reproduces reasonably well the main characteristics of the transverse
momentum distributions of kaons, protons and $\Lambda$'s.
The resulting correlation functions for all these particles are
of Gaussian form over a wide range of $Q_{\perp}^{2}$ and  $Q_{\parallel}^{2}$
(up to several hundred MeV) and thus the source radii are well defined.

The mass dependence of the calculated $R_{\parallel}$ and $R_{\perp}$ is
plotted in in Fig.~\ref{R_massdep}(b) where also the available results of 
the LEP experiments \cite{LEP_1d} are shown.

One sees that the inequality $R_{\parallel} >R_{\perp}$ is still
satisfied in the model, although the difference between the two radii at
higher masses is not as large as in the case of pions. 
The data points in this figure
correspond to the correlation radius $R_0$ determined in 1-dimensional
analyses (the only available data for heavy particles)\cite{LEP_1d}. 
Multiple entries of the result from the same experiment for pions correspond
to different measurements made with different reference samples.
The points at kaon mass represent measurements for both $K_{s}^{0}$
and $K^{\pm}$ pairs. The measurements for $\Lambda$ pairs come
from spin analysis (except for the second ALEPH point with small error)
where there is no need for a reference sample.  
The correspondence between $R_{0}$ and the two radii 
$R_{\parallel}$ and $R_{\perp}$ is not
obvious (at least experimentally), but the trend of the data is
reasonably well reproduced by the model. More accurate data on kaons
would be of great help to further elucidate this point.

\section{Conclusions and Comments}

In conclusion, we have found that the correlation between the momentum
and the production point of a produced hadron, suggested by the
Gottfried-Bjorken hypothesis of in-out cascade, seems to account
 (at least approximately) for the observed correlation between
identical particles observed in $e^+e^-$ annihilation. Together with
data on the single particle transverse momentum distribution, it predicts
strong anisotropy of the two-pion correlation function, in agreement
with the observations. The mass dependence of the "effective source radius" is
also adequately described. Large uncertainties, both in the theoretical
determination of the model parameters, and in the experimental data
do not allow, however, to obtain more quantitative conclusions.

Several comments are in order.

(i) One sees from the Table 1 that out of the three parameters
determined from transverse momentum distribution, only two ($v$ and 
$\delta_{\perp}$) are relatively well constrained, whereas
$\delta_{\parallel}$ is only weakly restricted. It is
interesting to note that - within the wide  range allowed - the value of 
$\delta_{\parallel}$ is consistent with the relation
$\delta_{\parallel}=\delta_{\perp}$ 
corresponding to the quasi-isotropic case.
 Accepting this 
isotropy, we see that the effective width  of the momentum distribution around 
the average given by the Bjorken-Gottfried condition (\ref{1}) is close
to 300 MeV, a rather reasonable value.

The small value of $v$ is also  recomforting, as it guarantees that
the transverse expansion of the "tube" from which the particles are
emitted is not unreasonably fast.

(ii) The mass dependence of the effective HBT correlation radii
calculated from the model and shown in Fig.6 was obtained under the
assumption that the parameters of the model do not depend on particle
masses. This assumption was verified (within large experimental
uncertainties) for $v$, $\delta_{\perp}$ and $\delta_{\parallel}$. 
No such direct check is available for $\tau_0$.
However, since the observed mass
dependence of the HBT correlation radii does agree -at least
approximately - with the experimental observations, we may take it
as an argument that {\it also $\tau_0$ does not depend on particle
masses}:
\be
\tau_0  \sim const(M_{\perp}).  \label{z1}
\ee
 This seems a rather non-trivial conclusion, as it indicates that
-within the Bjorken-Gottfried hypothesis (\ref{1})-  all particles are
emitted at, roughly, the same proper time $\tau_0$.

 This result may be
contrasted with a simple expectation (sometimes justified by uncertainty
principle) suggesting
\be
\tau_0 \sim \frac1{M_{\perp}}  \label{zz}
\ee
which would give a stronger drop of $R$ with increasing mass,
particularly for transverse radius.

It seems that also the models based on the picture of string decay 
\cite{models} would not fulfill the condition (\ref{z1}) but rather
obey
\be
\tau_0 \sim M_{\perp}.  \label{zz1}
\ee
Such a relation  corresponds, in our model,  to a much weaker dependence of the
correlation radii on the particle masses and thus  an other mechanism would 
have to be invented to describe the data \cite{and}.

(iii) In the present paper, studying the two-particle distribution, we
considered -following the approach employed in experimental analyses
\cite{LEP_2d}-
 the boost invariant variable $Q_{\perp}^2$ and the variable 
$Q_{\parallel}^2$ evaluated in the longitudinal center-of-mass system.
Assuming boost invariance and azimuthal symmetry of the distributions,
one finds that a complete analysis would involve 4 variables. As it is 
convenient to choose them boost invariant, one could use for
instance the two transverse momenta $|p_{1\perp}| $ and $|p_{2\perp}| $
the relative azimuthal angle $\phi_1-\phi_2$ and the relative rapidity
$y_1-y_2$. It would be interesting to see the data analysed in this way.

(iv) Another interpretation of the experimentally observed HBT
parameters was given in \cite{ax}. The authors take the point of view
that the observed HBT radii do indeed correspond to the actual size of
the particle emission region which is thus strongly dependent on the
particle mass. They argue that this dependence may be understood from
the uncertainty principle. It may be worth to point out that this
approach is rather different from ours. In our description the
parameters characterizing the particle emission region are {\it mass
independent} and the observed change in the HBT radii comes {\it solely}
from the momentum-position correlation as expressed in the 
assumed Bjorken-Gottfried condition (\ref{1}). 

(v) Our prescription  for construction of the two-particle correlation
function out of the single-particle source function is similar to -but not
identical with-  that advocated recently by Heinz and collaborators \cite{wh,uh}. 
The 
difference is in the treatment of the fourth component of the average momentum 
vector of the two particles [$\vec{P} \equiv (\vec{q}_1+\vec{q}_2)/2$]. 
As seen from (\ref{3.3b}), we use 
\be
P_0 \equiv \frac12 (E_1+E_2)    \label{z3}
\ee
which guarantees the four-vector character of $P_{\mu}$ but brings it
off mass-shell. Heinz et al. propose to take
\be
P_0= \sqrt{m^2+(\vec{P})^2}    \label{z4}
\ee
which insures $P_{\mu}P^{\mu}= m^2$ but induces complicated
transformation properties. In practice, however, one is only 
interested in the region
$\vec{q}_1 \approx \vec{q}_2$ where these two prescriptions are not
substantially different (the source functions themselves, however, 
are not the same).

%
%

\vspace{0.3cm}
{\bf Acknowledgements}
\vspace{0.3cm}

We would like to thank C.I.Tan, 
A.Krzywicki, W.Ochs and L.Stodolsky for interesting
discussions. 
This investigation was supported in part by  the Subsydium FNP 1/99 and  by the
KBN Grants: No 2 P03B 086 14, No 620/E-77/SPUB/CERN/P-03/DZ 3/99.

\end{document}